\PassOptionsToPackage{unicode}{hyperref}
\PassOptionsToPackage{hyphens}{url}
\documentclass[
]{article}
\usepackage{xcolor}
\usepackage[margin=1in]{geometry}
\usepackage{amsmath,amssymb}
\usepackage{iftex}
\ifPDFTeX
  \usepackage[T1]{fontenc}
  \usepackage[utf8]{inputenc}
  \usepackage{textcomp} 
\else 
  \usepackage{unicode-math} 
  \defaultfontfeatures{Scale=MatchLowercase}
  \defaultfontfeatures[\rmfamily]{Ligatures=TeX,Scale=1}
\fi
\usepackage{lmodern}
\ifPDFTeX\else
\fi
\IfFileExists{upquote.sty}{\usepackage{upquote}}{}
\IfFileExists{microtype.sty}{
  \usepackage[]{microtype}
  \UseMicrotypeSet[protrusion]{basicmath} 
}{}
\makeatletter
\@ifundefined{KOMAClassName}{
  \IfFileExists{parskip.sty}{%
    \usepackage{parskip}
  }{
    \setlength{\parindent}{0pt}
    \setlength{\parskip}{6pt plus 2pt minus 1pt}}
}{
  \KOMAoptions{parskip=half}}
\makeatother
\usepackage{graphicx}
\makeatletter
\newsavebox\pandoc@box
\newcommand*\pandocbounded[1]{
  \sbox\pandoc@box{#1}%
  \Gscale@div\@tempa{\textheight}{\dimexpr\ht\pandoc@box+\dp\pandoc@box\relax}%
  \Gscale@div\@tempb{\linewidth}{\wd\pandoc@box}%
  \ifdim\@tempb\p@<\@tempa\p@\let\@tempa\@tempb\fi
  \ifdim\@tempa\p@<\p@\scalebox{\@tempa}{\usebox\pandoc@box}%
  \else\usebox{\pandoc@box}%
  \fi%
}
\def\fps@figure{htbp}
\makeatother
\usepackage[]{natbib}
\usepackage{booktabs,longtable,array,multirow,colortbl,threeparttable,graphicx,float}
\usepackage[margin=1in]{geometry}

\usepackage{booktabs}
\usepackage{longtable}
\usepackage{array}
\usepackage{multirow}
\usepackage{wrapfig}
\usepackage{float}
\usepackage{colortbl}
\usepackage{pdflscape}
\usepackage{tabu}
\usepackage{threeparttable}
\usepackage{threeparttablex}
\usepackage[normalem]{ulem}
\usepackage{makecell}
\usepackage{xcolor}
\usepackage{bookmark}
\IfFileExists{xurl.sty}{\usepackage{xurl}}{} 
\makeatletter
\@ifundefined{xmpquote}{}{}
\makeatother
\hypersetup{
  pdftitle={When Does Survey-Aware Cross-Validation Matter? The ICC{,} Not the Design Effect},
  pdfauthor={M. Ehsan Karim and Md Belal Hossain},
  hidelinks,
  pdfcreator={LaTeX via pandoc}}

\title{When Does Survey-Aware Cross-Validation Matter? The ICC, Not the
Design Effect}
\author{M. Ehsan Karim and Md Belal Hossain\footnote{School of
  Population and Public Health, University of British Columbia; Centre
  for Advancing Health Outcomes, University of British Columbia,
  Vancouver, BC, Canada. Correspondence: M. Ehsan Karim,
  \href{mailto:ehsan.karim@ubc.ca}{\nolinkurl{ehsan.karim@ubc.ca}}}}
\date{}

\begin{document}
\maketitle
\begin{abstract}
K-fold cross-validation assumes exchangeable observations, violated by
the stratification, clustering, and unequal weights of complex sample
surveys. Design-respecting ``survey CV'' exists, but the question of
when the extra care changes any conclusion has remained open. We answer
it before validating any model, with two inexpensive diagnostics: the
within-cluster intraclass correlation (ICC) of the outcome and of a
preliminary linear predictor. A simulated positive control demonstrates
their sensitivity, with naive cross-validation growing optimistic about
new-cluster performance as the ICC rises while cluster-level folds stay
honest. We then evaluate paired naive-versus-design-respecting
cross-validation in three national health surveys (chronic-pain,
diabetes, and adolescent-suicidality prediction; penalized and
random-forest learners, plus an unpenalized comparator) --- one
reanalysis, one prospective application, and one prespecified screen. In
all three the diagnostics correctly anticipated the outcome: no scheme
difference of practical size, and the only interval excluding zero
showed pessimism, not the optimism that cluster leakage produces ---
even where the design effect was large (which, unlike the ICC, is not
the right trigger). We also show the stratified recipe is often
infeasible in public-use designs and give a fallback hierarchy, and we
document weight-handling errors whose order-of-magnitude artifacts
dwarfed any fold-scheme effect. Reproducible code accompanies the paper.
\end{abstract}


\section{Introduction}\label{introduction}

Predictive modeling with complex sample survey data is increasingly
common, and its validation machinery --- most prominently K-fold
cross-validation (CV) --- presumes exchangeable observations, an
assumption violated by the stratification, clustering, and unequal
selection probabilities of surveys such as the National Health Interview
Survey (NHIS) \citep{skinner2017introduction}. \cite{wieczorek2022kfold}
formalize the fix: choose folds and compute test errors in ways that
respect the design (``Survey CV''), implemented in the \texttt{surveyCV}
package \citep{guerin2022surveycv}. Related design-aware tools include
design-based prediction-error estimation \citep{holbrook2020estimating},
design-based information criteria \citep{lumley2015aic}, CV debiasing
under correlated data \citep{rabinowicz2022cross}, and replicate-weight
methods for penalized selection and ROC estimation
\citep{iparragirre2023variable, iparragirre2023estimation}. Yet
Wieczorek et al.~flag as an open question \emph{when} the
naive-versus-survey-aware discrepancy is large enough to matter ---
clustering makes naive CV overconfident, stratification acts in the
opposite direction, and the two can offset. Where Wieczorek et
al.~supply the estimator and \citet{rabinowicz2022cross} debias
correlated-data CV after the fact, we contribute an \emph{up-front},
survey-specific diagnostic --- empirically calibrated and paired with
feasibility guidance and weight-handling pitfalls --- that answers
whether the design-respecting machinery is needed at all before it is
run.

This article answers that question operationally, motivated by a
concrete case: an NHIS prediction study of high-impact chronic pain
\citep[HICP;][]{falasinnu2023problem, dahlhamer2018prevalence} that used
LASSO \citep{tibshirani1996regression} and random forest
\citep{breiman2001random} with random-fold CV --- an analysis the
present authors participated in, and one whose random-fold choice
remains routine in applied work (we note, owning the overlap, that
\citealt{wieczorek2022kfold} had appeared several months before that
study was submitted but had not yet reached applied practice). We make
three contributions. \textbf{First, a calibrated pre-analysis
diagnostic:} the within-cluster intraclass correlation (ICC) of the
outcome and of a preliminary linear predictor gauges the information a
naive fold split could leak; a simulated positive control shows the
diagnostic rising exactly when naive CV misleads --- and that the design
effect (DEFF), the familiar survey summary, is \emph{not} the right
trigger. \textbf{Second, a paired three-survey evaluation:} reanalyzing
the NHIS study (three learners, m = 20 imputations), then applying the
workflow prospectively to NHANES (diabetes) and, with a prespecified
decision rule, to the multistage, school-based Youth Risk Behavior
Survey (YRBS; suicidality), we find the diagnostics correctly predicted
a null scheme effect in all three. \textbf{Third, a practitioner's field
guide:} the textbook stratified Survey CV recipe is often
\emph{infeasible} in public-use designs --- we give a fallback hierarchy
--- and two weight-handling plug-in errors we document produced
order-of-magnitude artifacts that dwarfed any fold-scheme effect.

\section{Methods}\label{methods}

\subsection{Design diagnostics}\label{design-diagnostics}

For a survey outcome \(y\) with weights \(w\), strata, and
primary-sampling-unit (PSU) clusters --- the survey's first-stage
clusters --- we compute: (a) the design effect of the outcome mean ---
the \emph{design-based} ratio of the survey variance to the
simple-random-sampling variance (from \texttt{survey}, so it already
reflects weights, clustering, stratification, and their interactions)
--- which we descriptively factor as DEFF\(_{\text{total}} \approx\)
DEFF\(_{\text{weights}}\) \(\times\) DEFF\(_{\text{cluster+stratum}}\),
with DEFF\(_{\text{weights}} = 1 + \mathrm{CV}(w)^2\) Kish's
weights-only factor \citep{kish1965survey} and the residual quotient a
rough clustering-net-of-stratification summary; this
\emph{factorization} (not the design-based total) is exact only when
weights are unrelated to the design structure \citep{kish1992weighting}
--- which the informative oversampling in NHIS and NHANES violates, so
we read the split as illustrative only; and (b) one-way ANOVA intraclass
correlations within PSU-clusters, for the outcome and for the linear
predictor of a preliminary model fit to the full sample. Concretely: fit
a weighted LASSO once to the full sample (penalty at the
minimum-CV-error \(\lambda\)), extract its in-sample linear predictor
(on the logit scale), and compute unweighted one-way ANOVA ICCs of that
predictor and of the outcome within globally unique strata-by-PSU
clusters. Throughout, this single weighted-LASSO linear predictor
supplies the diagnostic ICC for \emph{all} learners, including the
random forest. The linear-predictor ICC gauges how much cluster-specific
signal a naive fold split could pass between training and test sets ---
the leakage mechanism by which naive CV becomes overoptimistic under
clustering \citep{wieczorek2022kfold, saeb2017need, kaufman2012leakage}.
It is an in-sample heuristic, not a bound, so we calibrate it with the
simulation below. Because a linear predictor is additive, it cannot
represent the cluster-specific non-linear interactions a random forest
could capture and leak; the linear-predictor ICC may therefore
\emph{understate} the clustering a tree learner could exploit. We
backstop it two ways: with the learner-independent \emph{outcome} ICC
--- an upper bound on the between-cluster signal in \(y\) available to
any learner, and here even lower (0.002--0.012) than the
linear-predictor value --- and, for the random forest, with the direct
naive-versus-design-respecting contrast reported in Table 1. A large
excess of the outcome ICC over the linear-predictor ICC would prompt
such a learner-native check. Our thesis, tested throughout: \emph{the
ICC, not the DEFF, is the diagnostic for the cluster-leakage optimism
that survey-aware folds guard against} --- stratification and unequal
weights shift naive CV in the opposite, conservative direction.

\subsection{Common evaluation
protocol}\label{common-evaluation-protocol}

We compare the two fold schemes as tools for \emph{internal validation}
--- estimating a fitted model's out-of-sample performance from the
sample at hand, meaning generalization to the rest of the finite
population beyond the sampled PSUs (the same new-cluster target the
positive control in \S2.4 makes explicit: design-respecting folds
estimate it by holding out whole PSUs, whereas naive folds leak
within-PSU signal). Every comparison is \emph{paired}: naive (seeded
random) and design-respecting folds are evaluated on identical data ---
the same 20 multiply imputed NHIS datasets
\citep{rubin1987multiple, vanbuuren2011mice}, or the same 10 seeded
repeats in NHANES and YRBS --- and we report the paired per-imputation
(or per-repeat) difference in AUC, the primary metric, with a
\(t\)-based 95\% CI (secondary metrics are shown per scheme);
overlapping marginal intervals are not a valid test for correlated
estimates. For NHIS the pairing is over the 20 imputations. For NHANES
and YRBS, whose repeats re-randomize folds on one fixed dataset, the
estimand is the expected scheme difference for that dataset over fold
randomization, and the intervals are Monte Carlo 95\% CIs whose widths
shrink with the number of repeats --- hence our emphasis on the
prespecified equivalence margin rather than zero-exclusion. Performance
metrics are survey-weighted throughout: weighted AUC (analytic variance
via the Hanley--McNeil formula with Kish effective sample sizes,
benchmarked against fully design-based replicate-weight standard errors
--- Web Appendix A), weighted Brier score, and a design-based
calibration slope. Multiple-imputation results pool by Rubin's rules
with the within-imputation variance taken as the per-fold variance
divided by the fold count. Full formulas, truncation and lonely-PSU
handling, and all scheme definitions are in Web Appendix A. Throughout,
\emph{naive CV} means seeded random folds that ignore the design in fold
\emph{assignment} (the survey weights still enter fitting and evaluation
in every scheme); \emph{stratified Survey CV} allocates PSUs to folds
within strata (\texttt{surveyCV::folds.svy}); \emph{PSU-level CV}
assigns whole PSUs to folds while ignoring strata; and
\emph{design-respecting} refers to either of the latter two.

\subsection{Three surveys}\label{three-surveys}

\textbf{NHIS 2016 (reanalysis).} HICP prediction on 32,980 adults; 43
predictors as in \cite{falasinnu2023problem}; m = 20 multiple
imputations (17.5\% of respondents had a missing predictor, imputed as
in that study; each scheme comparison is paired within imputation).
Learners: LASSO
\citep[nested 5-fold selection of $\lambda$ by the minimum-CV-error rule;][]{friedman2010regularization},
unpenalized survey-weighted logistic regression (\texttt{svyglm}), and a
500-tree probability random forest \citep{wright2017ranger} with survey
weights as case weights (an approximation --- case weights tilt only the
bootstrap draws; Web Appendix A). Schemes: naive 5-fold versus
survey-aware 5-fold (\texttt{folds.svy}, PSUs within strata), plus a
replicate-weight variance benchmark (Web Appendix A). Because the fold
scheme is the object of study, \(\lambda\) is selected by the same
procedure (naive inner CV) under both schemes, so differential tuning
cannot drive the contrast; a fully design-respecting pipeline would also
stratify the inner tuning folds (Web Appendix A.2).

\textbf{NHANES 2017--2018 (prospective).} Doctor-diagnosed diabetes; 14
predictors; complete case, n = 4,152 \citep{dinh2019data}. The public
design (15 strata \(\times\) 2 PSUs) makes stratified 5-fold Survey CV
\emph{infeasible} --- a \emph{per-stratum} limit (two PSUs per stratum)
that pooling additional NHANES cycles would not relax, since each added
cycle brings new strata rather than more PSUs per stratum --- so we
compare naive 5-fold, PSU-level 5-fold, stratified 2-fold, and a naive
2-fold learning-curve control (Web Appendix B).

\textbf{YRBS 2023 (prespecified prospective screen).} Considered-suicide
prediction on 9,838 high-school students; 16 predictors following
published YRBS-family models \citep{kim2024machine, lim2022prediction}.
The public-use design releases 16 strata and 68 primary sampling units
--- 73 unique strata-by-PSU clusters, with individual schools masked ---
so homogeneity is measured within these design clusters. The decision
rule was fixed \emph{before} any computation: the comparison would run
regardless, but a noticeable difference would be deemed plausible only
if the within-cluster linear-predictor ICC reached \(\approx\) 0.10.
Diagnostics were computed, the verdict recorded, and only then the
comparison run (naive versus PSU-level 5-fold; stratified folds are
infeasible for any \(K \geq 2\) because at least one stratum holds a
single PSU-cluster; Web Appendix C).

\subsection{A simulated positive
control}\label{a-simulated-positive-control}

Because all three surveys turn out to sit in the low-homogeneity regime,
they demonstrate only the diagnostics' specificity. For sensitivity, we
simulate finite populations of 400 clusters \(\times\) 60 units with a
binary outcome from a logistic model in 8 covariates; a single mixing
parameter \(\kappa \in \{0, 0.1, 0.2, 0.4, 0.6\}\) sets both the
cluster-level share of covariate variance and the scale of a cluster
random intercept (\(\sigma_u = 1.5\sqrt{\kappa}\)), so raising
\(\kappa\) raises the within-cluster ICC. For each of 200 replicate
samples per \(\kappa\) (60 whole clusters; \(n = 3{,}600\); equal
weights, since weighting is exercised by the empirical examples), an
unpenalized logistic model is validated by naive random 5-fold CV and by
cluster-level 5-fold CV, and both are compared with the model's true AUC
on the \emph{unsampled} clusters --- the estimand for generalization to
new clusters. The ICC diagnostics use the same one-way ANOVA estimator
as the survey analyses, with the fitted logistic linear predictor as the
in-sample proxy. We run this control in a base configuration (Figure 1)
and then \emph{factorially} --- three learners (logistic, LASSO, random
forest), higher discrimination (true AUC to 0.81), informative unequal
survey weights, a stratified design, unequal cluster sizes, and the
covariate- and intercept-clustering channels decoupled (Figure 2; the
full 84-condition grid and numbers are in Web Appendix D). All analyses
use R 4.5.1; seeded scripts reproducing every exhibit accompany the
paper (see Data and code availability).

\section{Results}\label{results}

\subsection{Calibration: the diagnostics flag real
danger}\label{calibration-the-diagnostics-flag-real-danger}

\begin{figure}
\centering
\pandocbounded{\includegraphics[keepaspectratio,alt={The positive control calibrates the diagnostic (base scenario, unpenalized logistic learner --- one cell of the Figure 2 grid). Mean bias of the CV estimate for new-cluster AUC (200 replicates per condition) as within-cluster homogeneity rises: naive random folds grow optimistic, cluster-level folds stay honest. Shaded band: the range of linear-predictor ICCs observed in NHIS, NHANES, and YRBS (0.020--0.043), all deep in the null regime. Figure 2 generalizes the naive-fold curve across learners and designs.}]{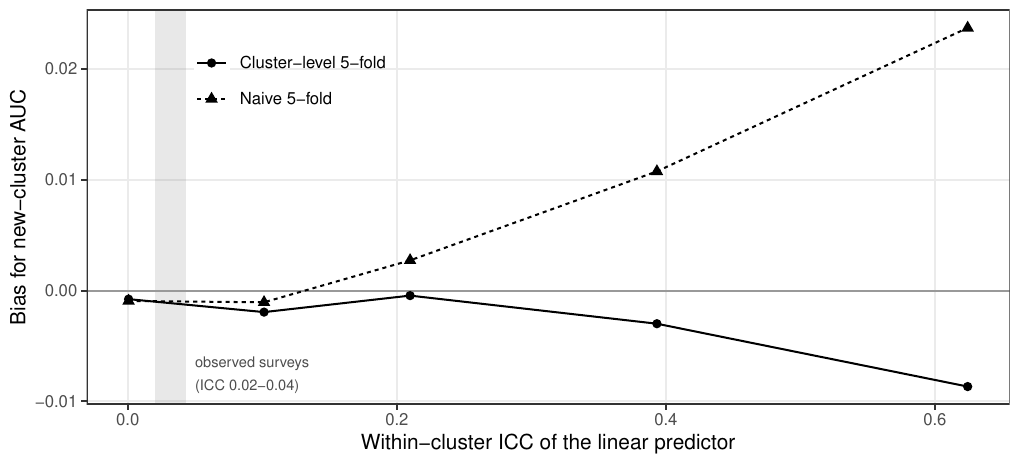}}
\caption{The positive control calibrates the diagnostic (base scenario,
unpenalized logistic learner --- one cell of the Figure 2 grid). Mean
bias of the CV estimate for new-cluster AUC (200 replicates per
condition) as within-cluster homogeneity rises: naive random folds grow
optimistic, cluster-level folds stay honest. Shaded band: the range of
linear-predictor ICCs observed in NHIS, NHANES, and YRBS (0.020--0.043),
all deep in the null regime. Figure 2 generalizes the naive-fold curve
across learners and designs.}
\end{figure}

Figure 1 establishes the ruler. With no cluster structure, both schemes
are unbiased for new-cluster AUC. As the linear-predictor ICC rises,
naive random-fold CV becomes progressively optimistic (bias +0.011 at
ICC \(\approx\) 0.39; +0.024 at 0.62) while cluster-level CV remains
approximately honest, and the outcome ICC rises in step (Web Appendix
D). An analyst computing these diagnostics before choosing a validation
strategy is warned as the danger emerges --- at least in this coupled,
linear-model scenario --- and can see where a given survey sits on the
ruler. In this calibration the naive-fold optimism is negligible at
\(\mathrm{ICC}_{lp} \approx 0.10\) (\(-0.001\)), reaches \(+0.003\) by
0.21 and \(+0.011\) by 0.39 (Web Appendix D); we therefore treat
\(\mathrm{ICC}_{lp} \gtrsim 0.1\) as a conservative trigger for
insisting on design-respecting folds --- the prespecified YRBS rule
(Section 2.3).

\begin{figure}
\centering
\pandocbounded{\includegraphics[keepaspectratio,alt={The diagnostic generalizes across learners and designs. Naive-CV optimism (mean naive-CV AUC minus the true new-cluster AUC; 200 replicates per condition) against the linear-predictor ICC, by learner (panels) and data-generating scenario. In the five designs where clustering enters the outcome (baseline, higher AUC to 0.81, unequal survey weights, stratified design, cluster-size imbalance) optimism rises with the ICC for every learner --- the random forest most --- crossing the 0.01 material-bias margin (dashed horizontal) near ICC 0.2--0.4. The covariate-only control isolates the mechanism: covariate clustering alone yields a high linear-predictor ICC with negligible optimism --- a conservative over-flag. (The random-intercept control, whose linear-predictor ICC stays \textbackslash approx 0 at all \textbackslash kappa, collapses onto the origin and is omitted; Table S7.) Per-scenario values: Web Appendix D, Table S7.}]{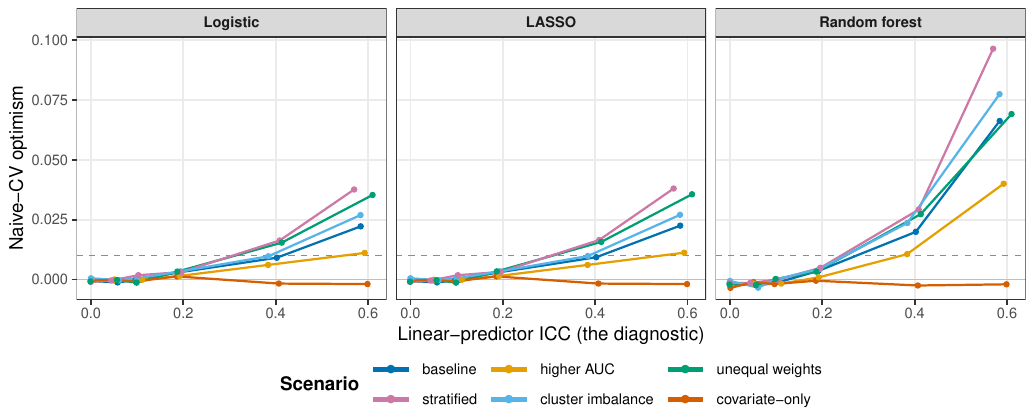}}
\caption{The diagnostic generalizes across learners and designs.
Naive-CV optimism (mean naive-CV AUC minus the true new-cluster AUC; 200
replicates per condition) against the linear-predictor ICC, by learner
(panels) and data-generating scenario. In the five designs where
clustering enters the outcome (baseline, higher AUC to 0.81, unequal
survey weights, stratified design, cluster-size imbalance) optimism
rises with the ICC for every learner --- the random forest most ---
crossing the 0.01 material-bias margin (dashed horizontal) near ICC
0.2--0.4. The covariate-only control isolates the mechanism: covariate
clustering alone yields a high linear-predictor ICC with negligible
optimism --- a conservative over-flag. (The random-intercept control,
whose linear-predictor ICC stays \(\approx 0\) at all \(\kappa\),
collapses onto the origin and is omitted; Table S7.) Per-scenario
values: Web Appendix D, Table S7.}
\end{figure}

Figure 2 shows the ruler is not an artifact of that single base,
logistic-learner scenario. Across the factorial --- adding a LASSO and a
random forest, raising discrimination to true AUC 0.81, and introducing
informative survey weights, a stratified design, and unequal cluster
sizes --- naive-CV optimism rises monotonically with the
linear-predictor ICC in every design where clustering enters the
outcome, for all three learners, with material bias (\(>0.01\)) again
emerging near \(\mathrm{ICC}_{lp}\) 0.2--0.4; the \(\gtrsim 0.1\)
trigger stays conservative throughout, and the random forest is the
\emph{most} affected learner, not an exception. Decoupling the two
clustering channels locates the mechanism: a pure cluster random
intercept leaves the linear-predictor ICC near zero and produces no
optimism (the outcome ICC alone would over-flag it), whereas covariate
clustering alone drives the linear-predictor ICC high yet yields
negligible optimism --- an over-flag in the safe direction. Leakage
tracks the predictable, outcome-relevant cluster signal the
linear-predictor ICC is built to capture.

\subsection{Three surveys, three predicted
nulls}\label{three-surveys-three-predicted-nulls}

\begin{table}[!h]
\centering
\caption{\label{tab:master-table}Design diagnostics and the paired naive-minus-design-respecting AUC difference across three national surveys and two learners; every paired difference falls well inside the prespecified equivalence margin $|\Delta\mathrm{AUC}| < 0.01$, and the only interval excluding zero lies in the pessimistic direction, opposite to the optimism cluster leakage would produce. \emph{Design}: number of strata / clustering units (the fold unit): NHIS, 655 stratum-by-PSU clusters from 111 recycled codes across 52 strata; NHANES, 30 PSUs ($=15\times2$); YRBS, 73 clusters (schools masked). \emph{DEFF}: design-based total design effect (from \texttt{survey}); the displayed weights $\times$ cluster-net-of-stratification split is an illustrative Kish decomposition, not exact when weights correlate with the design as in these oversampling surveys (components rounded independently, so their product may differ from the total in the last digit). \emph{ICC}: outcome / linear-predictor intraclass correlation, within clusters. \emph{$\Delta$AUC}: naive minus design-respecting (PSU-level) 5-fold AUC, with 95\% CI; NHIS pairs 20 imputations, NHANES and YRBS pair 10 seeded repeats (Monte Carlo CIs). \emph{Max strat.\ $K$}: largest fully stratified Survey-CV fold count the public design supports. Full per-scheme performance tables: Web Appendices A--C.}
\centering
\fontsize{8}{10}\selectfont
\begin{tabular}[t]{lllllll}
\toprule
\cellcolor[HTML]{E8ECF0}{\textbf{Survey}} & \cellcolor[HTML]{E8ECF0}{\textbf{Learner}} & \cellcolor[HTML]{E8ECF0}{\textbf{Design}} & \cellcolor[HTML]{E8ECF0}{\textbf{DEFF}} & \cellcolor[HTML]{E8ECF0}{\textbf{ICC}} & \cellcolor[HTML]{E8ECF0}{\textbf{$\Delta$AUC (95\% CI)}} & \cellcolor[HTML]{E8ECF0}{\textbf{Max strat.\ $K$}}\\
\midrule
 & LASSO &  &  &  & 0.0000 (-0.0003, 0.0003) & \\

\multirow{-2}{*}{\raggedright\arraybackslash NHIS 2016} & RF & \multirow{-2}{*}{\raggedright\arraybackslash 52 str.\ /\ 655 clus.} & \multirow{-2}{*}{\raggedright\arraybackslash 1.69 = 1.40$\times$1.20} & \multirow{-2}{*}{\raggedright\arraybackslash 0.012 / 0.043} & -0.0002 (-0.0007, 0.0003) & \multirow{-2}{*}{\raggedright\arraybackslash 5-fold}\\
\cmidrule{1-7}
 & LASSO &  &  &  & -0.0016 (-0.0044, 0.0011) & \\

\multirow{-2}{*}{\raggedright\arraybackslash NHANES 2017--18} & RF & \multirow{-2}{*}{\raggedright\arraybackslash 15 str.\ /\ 30 PSU} & \multirow{-2}{*}{\raggedright\arraybackslash 1.56 = 2.30$\times$0.68} & \multirow{-2}{*}{\raggedright\arraybackslash 0.002 / 0.020} & -0.0038 (-0.0074, -0.0002) & \multirow{-2}{*}{\raggedright\arraybackslash $K\!\le\!2$}\\
\cmidrule{1-7}
 & LASSO &  &  &  & 0.0005 (-0.0009, 0.0019) & \\

\multirow{-2}{*}{\raggedright\arraybackslash YRBS 2023} & RF & \multirow{-2}{*}{\raggedright\arraybackslash 16 str.\ /\ 73 clus.} & \multirow{-2}{*}{\raggedright\arraybackslash 18.4 = 2.12$\times$8.7} & \multirow{-2}{*}{\raggedright\arraybackslash 0.012 / 0.033} & -0.0002 (-0.0016, 0.0012) & \multirow{-2}{*}{\raggedright\arraybackslash none}\\
\bottomrule
\end{tabular}
\end{table}

Table 1 carries the empirical result. All three surveys sit at
linear-predictor ICCs of 0.02--0.04 --- an order of magnitude below
where Figure 1 shows naive CV beginning to mislead --- and, as the
diagnostics indicated (retrospectively for NHIS, prospectively for
NHANES, by prespecified rule for YRBS), every paired scheme difference
is practically null against our prespecified equivalence margin of
\(|\Delta\mathrm{AUC}| < 0.01\) (motivated by Figure 1, where
decision-relevant bias first appears near \(+0.011\)): none exceeds
0.004 in absolute value, and five of six intervals include zero. Because
every paired 95\% CI lies entirely within \(\pm 0.01\), the comparison
is operationally two one-sided tests of practical equivalence at that
margin --- a criterion on the bias-magnitude axis, distinct from and
less conservative than the ICC \(\gtrsim 0.1\) trigger. The one
exception (NHANES random forest, \(-0.0038\), Monte Carlo 95\% CI
\(-0.0074\) to \(-0.0002\)) is in the \emph{pessimistic} direction ---
naive folds slightly under-estimating performance --- consistent with
NHANES's stratification-heavy design (15 strata \(\times\) 2 PSUs) and
its near-zero linear-predictor ICC (0.020), and opposite to the optimism
that cluster leakage would produce. These small differences are not
artifacts of insensitive machinery: discrimination, accuracy,
calibration, and selected model size are essentially identical across
schemes for every survey--learner pair, and for the flagship NHIS LASSO
case both cross-checks agree --- the unpenalized \texttt{svyglm}
comparator (point-estimate agreement) and a fully design-based
replicate-weight variance arm, whose standard error averages 0.79 times
the analytic value (Web Appendix A). Nor do they reflect low
information: the same reanalysis reproduces the published NHIS AUCs
\citep[0.891 LASSO, 0.885 random forest;][]{falasinnu2023problem} while
equipping them with design-based uncertainty (Web Appendix A).

Two further patterns matter. First, the DEFF column shows why it is the
wrong trigger: NHANES's clustering component is \emph{below} 1
(stratification over-compensates clustering --- the regime where naive
folds, if anything, overestimate error), while YRBS's residual component
looks alarming (8.7) purely because each of its large primary sampling
units is simultaneously the sampling and the weighting unit, inflating
the heuristic quotient; its ICCs are tiny. Had we triggered on DEFF we
would have expected a large YRBS effect; the ICC said otherwise, and the
ICC was right. Second, feasibility binds before optimality: stratified
Survey CV supports at most \(K = 2\) folds in NHANES and \emph{no}
stratified folds in YRBS (a stratum with a single PSU-cluster), and
within NHIS the fully stratified recipe fails for the two smallest
race/ethnicity subgroups (Web Appendix A). We therefore recommend an
explicit fallback hierarchy: (a) the largest feasible stratified \(K\);
then (b) PSU-level folds ignoring strata --- clustering is the leakage
channel, so this preserves the essential protection; then (c) reporting
design-respecting validation as not estimable, rather than silently
reverting to naive folds. In our examples every feasible rung gave the
same answer (the NHANES naive 2-fold control confirms the small
\(K = 2\) decrement is a learning-curve effect, not leakage; Web
Appendix B).

\subsection{The plug-in errors that did
matter}\label{the-plug-in-errors-that-did-matter}

While the fold choice proved immaterial, two weight-handling errors in
our own initial pipeline were anything but. Supplying raw NHIS weights
(mean \(\approx\) 3,200) as \texttt{glm} case weights in the calibration
regression drove the fit to quasi-complete separation and returned
slopes near \(10^{14}\) --- with no convergence warning; the corrected
design-based slope is 1.00. Feeding sums of weights (population counts)
instead of effective sample sizes into the Hanley--McNeil variance
formula shrank the AUC standard error \(\approx\) 67-fold, producing a
pooled interval roughly eight-fold too narrow (Rubin's
between-imputation variance, unaffected by the error, caps the
shrinkage) --- narrow enough to make trivial scheme differences look
meaningful, which it did in early drafts. Both artifacts, plus two
silent API traps (a scoring function that ignores its weight argument;
PSU codes that recycle across strata and silently merge clusters if
folds are built on raw codes), are documented with remedies in Web
Appendix A.6. The generic sanity checks: calibration slopes live on the
scale of 1; interval widths must be consistent with the \emph{effective}
sample size; and any analytic variance deserves one design-based
(replicate-weight) benchmark before it is reported.

\section{Discussion}\label{discussion}

\textbf{Summary.} Survey design must be respected in validation, and
\texttt{surveyCV} makes design-respecting folds one line of code --- but
across three national surveys, three outcomes, and two learners (plus an
unpenalized comparator), the choice of folds changed nothing of
practical consequence. That null is \emph{conditional}, not universal:
survey-aware folds matter when within-cluster homogeneity (the ICC) is
material, and all three surveys sat in the low-ICC regime where two
cheap diagnostics --- computed in advance for NHANES and YRBS,
retrospectively for NHIS --- correctly said they would not. The
practical dangers in this terrain were elsewhere: in feasibility
constraints the textbook recipe silently hits, and in the mechanical
insertion of population-scale weights into software that interprets them
as case counts.

\textbf{When should the choice matter?} The published ICC literature
maps the terrain without survey-by-survey fishing. Individual-level
health outcomes cluster weakly within the large geographic PSUs of
national surveys and within primary-care practices
\citep[median ICC 0.010 across 1,039 variables;][]{adams2004patterns, gulliford1999components}
--- the null regime of all three surveys here. The higher-ICC settings
are \emph{households and families} (ICCs up to about 0.3 for behavioral
and socioeconomic variables, well above the geographic-PSU values;
\citealt{gulliford1999components}), \emph{schools for academic outcomes}
(0.1--0.25; \citealt{hedges2007intraclass}) --- notably higher than the
PSU-clustered health-behavior ICCs we measured (individual schools are
masked in the public-use YRBS, so the accessible unit is the coarser
PSU) --- and \emph{village-clustered surveys} in low- and middle-income
countries (0.03--0.71; \citealt{dwivedi2022intra}); it is no accident
that the example where \cite{wieczorek2022kfold} saw Survey CV change
the selected model was a village-clustered LMIC survey. Analysts in
those settings should treat design-respecting folds as the default and
expect them to matter; analysts elsewhere can check in minutes. The
diagnostic complements rather than competes with post-hoc debiasing of
CV under correlated data \citep{rabinowicz2022cross} and
replicate-weight selection machinery \citep{iparragirre2023variable}: it
chooses honest folds a priori, at essentially zero cost, before any such
machinery is invoked.

\textbf{Limitations.} Our nulls are demonstrated, not universal: models
with strongly cluster-correlated predictors, or surveys with fewer and
more homogeneous clusters, can differ, and the diagnostics identify
exactly those cases. Public-use design variables are masked; current
Survey CV does not accommodate post-stratified or nonresponse-adjusted
weights \citep{wieczorek2022kfold}; the NHANES and YRBS examples are
complete-case methodological demonstrations; the random-forest weighting
is a resampling approximation whose probability calibration varied by
dataset (slope 1.10 in the NHIS, 0.85 in the smaller, higher-prevalence
YRBS) but was identical across fold schemes, leaving the
naive-versus-design-respecting contrast unaffected; and the base
simulation is stylized (equal cluster sizes, equal weights, one linear
learner, a single mixing parameter). A factorial extension (Web Appendix
D) relaxes each of these --- unequal weights and cluster sizes, a
stratified design, higher discrimination (true AUC to 0.81), a random
forest, and covariate- versus intercept-clustering decoupled --- and
\emph{demonstrates} that naive-CV optimism rises with the
linear-predictor ICC in every outcome-clustered design and learner (the
random forest included), with the \(\mathrm{ICC} \approx 0.1\) trigger
conservative throughout; the decoupling further shows the
linear-predictor ICC flags the predictable cluster signal (for a pure
random intercept it correctly reads near zero; for covariate-only
clustering it over-flags, the safe direction). The trigger remains a
calibrated screening threshold rather than a formally derived cutoff.
Our scope is cross-validatory \emph{internal} validation; resampling
alternatives --- the bootstrap and optimism-corrected (.632+) estimators
--- raise the same naive-versus-design-respecting question and are a
natural extension we do not pursue.

\textbf{A pragmatic workflow.} (1) Use the survey weights in fitting
\emph{and} evaluation, with effective-sample-size or design-based
uncertainty --- never raw weights in case-weight arguments or count
formulas. (2) Default to design-respecting folds
(\texttt{surveyCV::folds.svy}) for any learner. (3) Check feasibility
first (\(K\) stratified folds need \(K\) PSUs per stratum) and descend
the fallback hierarchy explicitly rather than reverting silently to
naive folds. (4) Compute the DEFF decomposition and the two within-PSU
ICCs before interpreting --- or assuming --- any scheme difference,
treating \(\mathrm{ICC}_{lp} \gtrsim 0.1\) as a conservative trigger for
insisting on design-respecting folds. (5) Compare schemes by paired
differences on shared folds, never by eyeballing marginal intervals. (6)
Sanity-check every metric's magnitude against its natural scale, ideally
against one replicate-weight benchmark, and report per established
guidance \citep{collins2015transparent}. Whether survey-aware
cross-validation will change a conclusion is checkable \emph{before} any
model is validated --- from the within-cluster ICC of a preliminary
model's predictions, not the design effect survey intuition reaches for
first; design-respecting folds are sometimes infeasible --- which is
worth reporting; and across three national health surveys the ICC
correctly anticipated that the folds would be immaterial, while two
mundane weight-handling errors could have reversed the study's
conclusions.

\section*{Data and code availability}\label{data-and-code-availability}
\addcontentsline{toc}{section}{Data and code availability}

NHIS 2016 (\url{https://www.cdc.gov/nchs/nhis/}), NHANES 2017--2018
(\url{https://wwwn.cdc.gov/nchs/nhanes/}), and YRBS 2023
(\url{https://www.cdc.gov/yrbs/data/}) are public CDC data; no
restricted data were used. Seeded scripts regenerating every exhibit,
and the Supporting Information (Web Appendices A--D), accompany the
paper at \url{https://github.com/ehsanx/survey-cv-diagnostics} (R 4.5.1
with \texttt{glmnet}, \texttt{survey}, \texttt{surveyCV},
\texttt{WeightedROC}, \texttt{mice}, \texttt{ranger}, and
\texttt{lme4}).

\section*{Acknowledgements}\label{acknowledgements}
\addcontentsline{toc}{section}{Acknowledgements}

This research was supported in part through computational resources from
Advanced Research Computing at the University of British Columbia.
During this work, the authors used AI-based tools (large language
models) to assist with the analysis and simulation code, and text
editing; the authors verified all outputs and take full responsibility
for the content.

\section*{Conflict of interest}\label{conflict-of-interest}
\addcontentsline{toc}{section}{Conflict of interest}

MEK and MBH were co-authors of \citet{falasinnu2023problem}, the NHIS
chronic-pain prediction study that provides this paper's NHIS example.

\section*{Funding}\label{funding}
\addcontentsline{toc}{section}{Funding}

NSERC Discovery Grant (PG\#: 20R01603) and Discovery Launch Supplement
(PG\#: 20R12709).

\section*{Supplementary Material}\label{supplementary-material}
\addcontentsline{toc}{section}{Supplementary Material}

Web Appendices A--D, Web Tables S1--S6, and Web Figure S1 are provided
in the Supplementary Material accompanying this paper.

\bibliography{references}

\begin{thebibliography}{}

\bibitem[Adams et~al., 2004]{adams2004patterns}
Adams, G., Gulliford, M.~C., Ukoumunne, O.~C., Eldridge, S., Chinn, S., and Campbell, M.~J. (2004).
\newblock Patterns of intra-cluster correlation from primary care research to inform study design and analysis.
\newblock {\em Journal of Clinical Epidemiology}, 57(8):785--794.

\bibitem[Breiman, 2001]{breiman2001random}
Breiman, L. (2001).
\newblock Random forests.
\newblock {\em Machine Learning}, 45(1):5--32.

\bibitem[Collins et~al., 2015]{collins2015transparent}
Collins, G.~S., Reitsma, J.~B., Altman, D.~G., and Moons, K. G.~M. (2015).
\newblock Transparent reporting of a multivariable prediction model for individual prognosis or diagnosis ({TRIPOD}): the {TRIPOD} statement.
\newblock {\em BMC Medicine}, 13:1.

\bibitem[Dahlhamer et~al., 2018]{dahlhamer2018prevalence}
Dahlhamer, J., Lucas, J., Zelaya, C., Nahin, R., Mackey, S., DeBar, L., Kerns, R., Von~Korff, M., Porter, L., and Helmick, C. (2018).
\newblock Prevalence of chronic pain and high-impact chronic pain among adults --- {U}nited {S}tates, 2016.
\newblock {\em MMWR Morbidity and Mortality Weekly Report}, 67(36):1001--1006.

\bibitem[Dinh et~al., 2019]{dinh2019data}
Dinh, A., Miertschin, S., Young, A., and Mohanty, S.~D. (2019).
\newblock A data-driven approach to predicting diabetes and cardiovascular disease with machine learning.
\newblock {\em BMC Medical Informatics and Decision Making}, 19:211.

\bibitem[Dwivedi et~al., 2022]{dwivedi2022intra}
Dwivedi, L.~K., Mahapatra, B., Bansal, A., Gupta, J., Singh, A., and Roy, T.~K. (2022).
\newblock Intra-cluster correlations in socio-demographic variables and their implications: an analysis based on large-scale surveys in {I}ndia.
\newblock {\em SSM - Population Health}, 21:101317.

\bibitem[Falasinnu et~al., 2023]{falasinnu2023problem}
Falasinnu, T., Hossain, M.~B., Weber, II, K.~A., Helmick, C.~G., Karim, M.~E., and Mackey, S. (2023).
\newblock The problem of pain in the united states: A population-based characterization of biopsychosocial correlates of high impact chronic pain using the national health interview survey.
\newblock {\em The Journal of Pain}, 24(6):1094--1103.

\bibitem[Friedman et~al., 2010]{friedman2010regularization}
Friedman, J., Hastie, T., and Tibshirani, R. (2010).
\newblock Regularization paths for generalized linear models via coordinate descent.
\newblock {\em Journal of statistical software}, 33(1):1.

\bibitem[Guerin et~al., 2022]{guerin2022surveycv}
Guerin, C., McMahon, T., and Wieczorek, J. (2022).
\newblock {\em surveyCV: Cross validation based on survey design}.
\newblock R package version 0.1.1.

\bibitem[Gulliford et~al., 1999]{gulliford1999components}
Gulliford, M.~C., Ukoumunne, O.~C., and Chinn, S. (1999).
\newblock Components of variance and intraclass correlations for the design of community-based surveys and intervention studies: data from the {H}ealth {S}urvey for {E}ngland 1994.
\newblock {\em American Journal of Epidemiology}, 149(9):876--883.

\bibitem[Hedges and Hedberg, 2007]{hedges2007intraclass}
Hedges, L.~V. and Hedberg, E.~C. (2007).
\newblock Intraclass correlation values for planning group-randomized trials in education.
\newblock {\em Educational Evaluation and Policy Analysis}, 29(1):60--87.

\bibitem[Holbrook et~al., 2020]{holbrook2020estimating}
Holbrook, A., Lumley, T., and Gillen, D. (2020).
\newblock Estimating prediction error for complex samples.
\newblock {\em Canadian Journal of Statistics}, 48(2):204--221.

\bibitem[Iparragirre et~al., 2023a]{iparragirre2023estimation}
Iparragirre, A., Barrio, I., and Arostegui, I. (2023a).
\newblock Estimation of the {ROC} curve and the area under it with complex survey data.
\newblock {\em Stat}, 12(1):e635.

\bibitem[Iparragirre et~al., 2023b]{iparragirre2023variable}
Iparragirre, A., Lumley, T., Barrio, I., and Arostegui, I. (2023b).
\newblock Variable selection with {LASSO} regression for complex survey data.
\newblock {\em Stat}, 12(1):e578.

\bibitem[Kaufman et~al., 2012]{kaufman2012leakage}
Kaufman, S., Rosset, S., Perlich, C., and Stitelman, O. (2012).
\newblock Leakage in data mining: formulation, detection, and avoidance.
\newblock {\em ACM Transactions on Knowledge Discovery from Data}, 6(4):1--21.

\bibitem[Kim et~al., 2024]{kim2024machine}
Kim, H., Son, Y., Lee, H., Kang, J., Hammoodi, A., Choi, Y., Kim, H.~J., Lee, H., Fond, G., Boyer, L., Kwon, R., Woo, S., and Yon, D.~K. (2024).
\newblock Machine learning--based prediction of suicidal thinking in adolescents by derivation and validation in 3 independent worldwide cohorts: algorithm development and validation study.
\newblock {\em Journal of Medical Internet Research}, 26:e55913.

\bibitem[Kish, 1965]{kish1965survey}
Kish, L. (1965).
\newblock {\em Survey Sampling}.
\newblock John Wiley \& Sons, New York.

\bibitem[Kish, 1992]{kish1992weighting}
Kish, L. (1992).
\newblock Weighting for unequal {$P_i$}.
\newblock {\em Journal of Official Statistics}, 8(2):183--200.

\bibitem[Lim et~al., 2022]{lim2022prediction}
Lim, J.~S., Yang, C.-M., Baek, J.-W., Lee, S.-Y., and Kim, B.-N. (2022).
\newblock Prediction models for suicide attempts among adolescents using machine learning techniques.
\newblock {\em Clinical Psychopharmacology and Neuroscience}, 20(4):609--620.

\bibitem[Lumley and Scott, 2015]{lumley2015aic}
Lumley, T. and Scott, A. (2015).
\newblock {AIC} and {BIC} for modeling with complex survey data.
\newblock {\em Journal of Survey Statistics and Methodology}, 3(1):1--18.

\bibitem[Rabinowicz and Rosset, 2022]{rabinowicz2022cross}
Rabinowicz, A. and Rosset, S. (2022).
\newblock Cross-validation for correlated data.
\newblock {\em Journal of the American Statistical Association}, 117(538):718--731.

\bibitem[Rubin, 1987]{rubin1987multiple}
Rubin, D.~B. (1987).
\newblock {\em Multiple Imputation for Nonresponse in Surveys}.
\newblock John Wiley \& Sons, New York.

\bibitem[Saeb et~al., 2017]{saeb2017need}
Saeb, S., Lonini, L., Jayaraman, A., Mohr, D.~C., and Kording, K.~P. (2017).
\newblock The need to approximate the use-case in clinical machine learning.
\newblock {\em GigaScience}, 6(5):gix019.

\bibitem[Skinner and Wakefield, 2017]{skinner2017introduction}
Skinner, C. and Wakefield, J. (2017).
\newblock Introduction to the design and analysis of complex survey data.
\newblock {\em Statistical Science}, 32(2):165--175.

\bibitem[Tibshirani, 1996]{tibshirani1996regression}
Tibshirani, R. (1996).
\newblock Regression shrinkage and selection via the lasso.
\newblock {\em Journal of the Royal Statistical Society: Series B}, 58(1):267--288.

\bibitem[van Buuren and Groothuis-Oudshoorn, 2011]{vanbuuren2011mice}
van Buuren, S. and Groothuis-Oudshoorn, K. (2011).
\newblock mice: Multivariate imputation by chained equations in {R}.
\newblock {\em Journal of Statistical Software}, 45(3):1--67.

\bibitem[Wieczorek et~al., 2022]{wieczorek2022kfold}
Wieczorek, J., Guerin, C., and McMahon, T. (2022).
\newblock K-fold cross-validation for complex sample surveys.
\newblock {\em Stat}, 11(1):e454.

\bibitem[Wright and Ziegler, 2017]{wright2017ranger}
Wright, M.~N. and Ziegler, A. (2017).
\newblock ranger: a fast implementation of random forests for high dimensional data in {C++} and {R}.
\newblock {\em Journal of Statistical Software}, 77(1):1--17.

\end{thebibliography}

\end{document}